\documentclass[preprint]{elsarticle}

\usepackage{graphicx}

\usepackage{amsmath}
\usepackage{amssymb}

\usepackage[font=footnotesize,labelfont=bf]{caption}
\usepackage[font=footnotesize,labelfont=bf]{subcaption}

\usepackage{booktabs}

\journal{Journal of Network and Computer Applications}

\begin{document}

\begin{frontmatter}



\title{Efficient image deployment in Cloud environments}


\author{{\'A}lvaro L{\'o}pez Garc{\'i}a\corref{cor1}}

\cortext[cor1]{Corresponding author}
\ead{aloga@ifca.unican.es}

\author{Enol Fern{\'a}ndez del Castillo}
\ead{enolfc@ifca.unican.es}

\address{Instituto de F{\'i}sica de Cantabria --- IFCA (CSIC---UC).\\
Avda. los Castros s/n. 39005 Santander, Spain}

\begin{abstract}
The biggest overhead for the instantiation of a virtual machine in a cloud
infrastructure is the time spent in transferring the image of the virtual
machine into the physical node that executes it. This overhead becomes larger
for requests composed of several virtual machines to be started concurrently,
and the illusion of flexibility and elasticity usually associated with the
cloud computing model may vanish. This poses a problem for both the resource
providers and the software developers, since tackling those overheads is
not a trivial issue.

In this work we implement and evaluate several improvements for virtual
machine image distribution problem in a cloud infrastructure and propose
a method based on BitTorrent and local caching of the virtual machine images
that reduces the transfer time when large requests are made.
\end{abstract}

\begin{keyword}
Cloud Computing \sep Image Deployment \sep OpenStack \sep Scheduling


\end{keyword}

\end{frontmatter}

\section{Introduction}
\label{sec:introduction}

As it is widely known, the Cloud Computing model is aimed on delivering
resources (such as virtual machines, storage and network capacity) as an on
demand service.  The most accepted publication defining the Cloud from the
United States National Institute of Standards and Technology (NIST), emphasizes
the \emph{rapid elasticity} as one of the essential characteristics of the
Cloud Computing model: ``capabilities can be elastically provisioned and
released, (...), to scale rapidly outward and inward (...)''~\cite{Mell2011}.
Moreover, users and consumers consider them as the new key features that are
more attractive \cite{Zhang2010,Armbrust2010} when embracing the cloud.

If we take into consideration virtual machines delivered by an Infrastructure
as a Service resource provider, these two outstanding features imply two
different facts. On the one hand, \textbf{elasticity} is the ability to start
and dispose one or several virtual machines (VMs) almost immediately.  On the
other hand, an \textbf{on demand} access implies that VMs are allocated
whenever the user requires them, without prior advise and without human
intervention from the Resource Provider (RP).

Any cloud must be able to deliver rapidly the requested machines to provide a
satisfactory elastic and on-demand perception according to any Service Level
Agreement SLA \cite{Aceto2012} established with the users or customers. With
this fact in mind, the Cloud Management Frameworks (CMFs) ---and as a
consequence the resource providers operating a cloud--- face a challenge when
they are requested to provision a large number of resources, specially when
running large infrastructures \cite{ChenZ2009}.

These on-demand and elastic perceptions that a cloud should be able to deliver
mostly depend on the time needed to serve the final service, so a rapid
provisioning should be one of the objectives of any cloud provider. Besides the
delays introduced by the Cloud Management Framework there are other factors
contributing to this delay from the user standpoint. For instance,
inter-datacenters transfers of large amounts of data \cite{Femminella2014} are
a good example of contributors to the final delivery time. Any reduction in
each of these factors will yield on a better reactivity of the cloud, leading
to an increase of the ability to satisfy elastic requests on-demand.

Therefore, in order to deliver a rapid service, this spawning delay or penalty
has to be decreased. It is the duty of the cloud provider to be able to
provision efficiently the resources to the users, regardless of the size of the
request, minimizing the costs of mapping the request into the underlying
resources\cite{Manvi2014}. Hence, it is needed to study how current CMFs can
minimize the start time of the virtual machines requested.  Our main
contribution in this paper is the proposal of an improvement of the current
Coud Management Frameworks in two sides: firstly, the CMFs should implement more
advanced and appropriated image transfer mechanisms; secondly, the cloud
schedulers should be adapted so as to make use of local caches on the physical
nodes. Moreover, in this paper:

\begin{itemize}
    \item We will study how the deployment of images into the physical machines
        poses a problem to an Infrastructure as a Service (IaaS) resource
        provider and how it introduces a penalty towards the users.

    \item We will discuss several image transfer methods that alleviate this
        problem and review the related work.

    \item We implement and evaluate some of the described methods in an
        existing CMF.

    \item We propose an improvement of the scheduling algorithm to take profit
        of the VM images cached at the physical nodes.
\end{itemize}

The paper is structured as follows:
In Section~\ref{sec:problem} we discuss and present the problem statement that.
In Section~\ref{sec:related} the related research in the area is presented and
discussed.
Section~\ref{sec:evaluation} contains the evaluation of some of the methods
described in the previous section.
In Section~\ref{sec:cache scheduler} we propose a modification to the
scheduling algorithms, and evaluate in combination with the studied image
transfer methods.
Finally, conclusions and future works are outlined in
Section~\ref{sec:conclusions}.

\section{Problem statement}
\label{sec:problem}

Whenever a virtual machine is spawned its virtual image disk must be available
at the physical node in advance. If the image is not available on that host, it
needs to be transfered, therefore the spawning will be delayed until the
transfer is finished. This problem is specially magnified if the request
consist on more than a few virtual machines, as more data needs to be
transfered over the network. As the underlying infrastructure increases its
size the problem becomes also bigger, the number of requests need to be
satisfied may become larger.

Regardless of the Cloud Management Framework (CMF) being used, the process of
launching a VM in an IaaS cloud infrastructure comprises a set of common steps:

\begin{enumerate}
    \item A VM image is created by somebody ---e.g. by a system administrator
        or a seasoned user---, containing the desired software environment.
    \item The image is uploaded to the cloud infrastructure image catalog or
        image repository. This image is normally stored as read-only,
        therefore, if further modifications (for example any user
        customization) need to be done on a given image, a new one must
        be created.
    \item VMs based on this image are spawned into the physical machines.
    \item The running VMs are customized on boot time to satisfy the user
        needs. This step is normally referred as contextualization and it is
        performed by the users.
\end{enumerate}

The first two steps are normally performed once in the lifetime of a virtual
machine image, meaning that once the image is created and is available in
the catalog, then it is ready for being launched, so there is no need to
recreate the image and upload it again. Therefore, assuming that the IaaS
provider is able to satisfy the request (i.e. there are enough available
resources to execute the requested VMs), whenever a user launches a VM, only
the two former steps will introduce a delay in the boot time.

The last step, that is, the contextualization phase is made once the virtual
instance has booted, and it is normally a user's responsibility and beyond the
Cloud Management Framework control~\cite{Campos2013,Li2012}.
Hence, the field where a IaaS resource provider can take actions to reduce
the boot time of a virtual machine is the spawning phase. This phase involves
several management and preparation operations that will depend on the Cloud
Management Framework being used. Generally, these operations will
consist on one or several of the following steps:

\begin{description}
    \item[Scheduling phase] where the software selects the most suitable nodes
        to satisfy the user's request.
    \item[Image transfer] if the image data is not available in the selected
        physical machine, the CMF has to transfer it from the catalog into that
        host.
    \item[Image duplication] once the image is available at the node. Some
        CMF duplicate the image before spawning the virtual machine. This way,
        the original image remains intact and it can be reused afterwards for
        another VM based on that same image.
    \item[Image preparation] consisting in all the further image modifications
        prior to the virtual machine spawning, needed to satisfy the user's
        request.
        For instance, this step can comprise the image resize, image format
        conversion, user-data injection into the image, file system checks,
        etc.
\end{description}

\begin{figure}[ht!]
    \centering
    \includegraphics[width=1\linewidth]{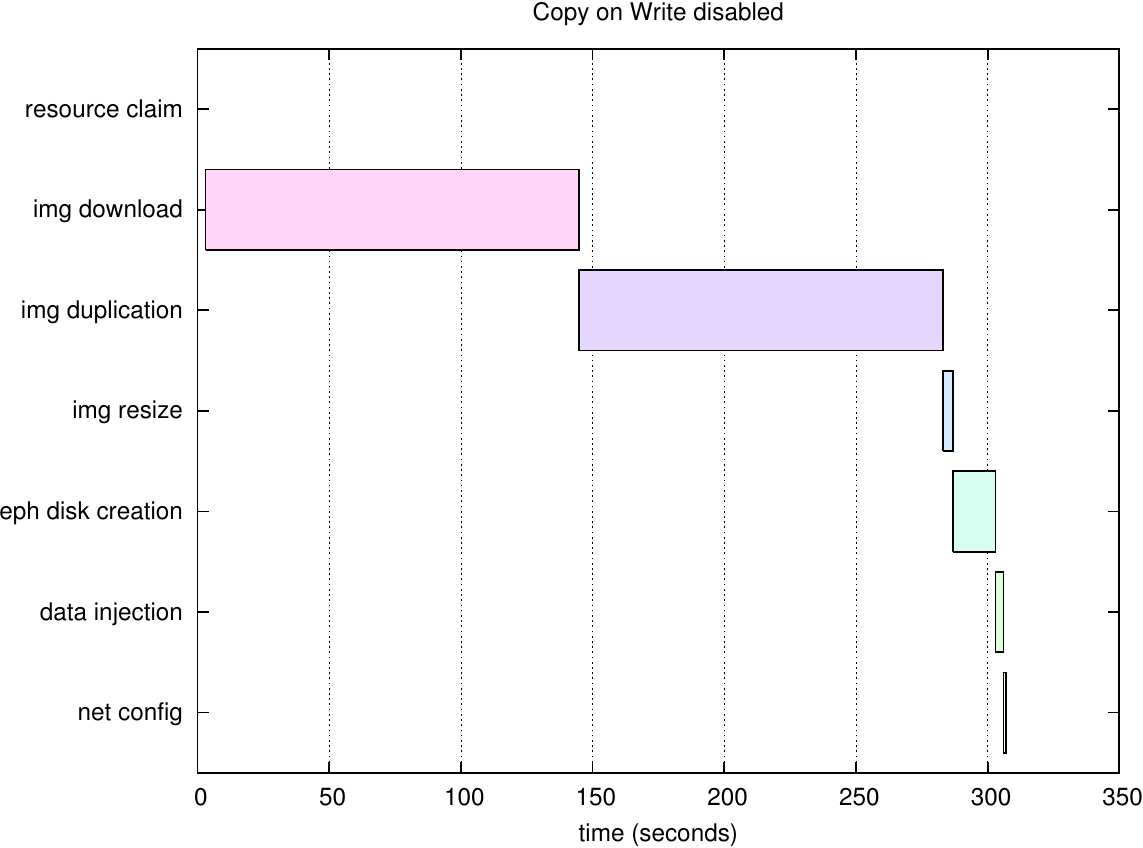}
    \caption{Chart of the boot process for one VM on an OpenStack cloud. The
    image used was 10GB large with an 80GB ephemeral disk.}
    \label{fig:bootchart-raw}
\end{figure}

Taking as an example the OpenStack cloud testbed described in the experimental
setup of Section~\ref{sec:testbed}, Figure~\ref{fig:bootchart-raw} shows the
boot sequence for an instance once the request is scheduled into a physical
machine. In this request a 10GB image was launched with an additional local
ephemeral disk of 80GB. This ephemeral empty space is created on the fly on the
local disk of the physical machine, therefore it is not transferred over the
network. In this initial setup, the images are stored in the catalog server and
are transferred using HTTP when they are needed in the compute host.

As it can be seen, the OpenStack spawning process is broken down into several
sub steps:

\begin{description}
    \item[Resource claim] The compute node checks if the requested resources
        are available, and claims them before spawning the instance.
    \item[Image download] The image is fetched from the image catalog, and it
        is stored in the local disk.
    \item[Image duplication] An exact replica image is created from the
        downloaded one.
    \item[Image resize] The image is resized to fit into the size request by
        the user. Normally minimal images are stored in order to spare disk and
        save transfer times, therefore these images need to be resized into
        the correct final size.
    \item[Ephemeral disk creation] An ephemeral virtual disk is created in the
        local disk. This virtual disk is created on the fly and it is normally
        located on the local machine disk, since it is a disposable space
        destroyed when the instance is terminated.
    \item[Data injection] Any data specified by the user is injected into the
        image. This step needs to figure out the image layout and try to inject
        the data into the correct location. This is a prone to errors step since
        the image structure is unknown to the middleware and therefore it can
        fail. It could be avoided with the usage of contextualization,
        assuming that the images are properly configured.
    \item[Network configuration] The virtual network is configured and set up
        in the physical node to ensure that the instance will have
        connectivity.
\end{description}

The \emph{Resource claim} step belongs to the \emph{Scheduling phase}, and the
steps labeled \emph{Image resize}, \emph{Ephemeral disk creation}, \emph{Data
injection}, \emph{Network configuration} belong to the aforementioned
\emph{Image preparation} phase. Observing Figure~\ref{fig:bootchart-raw} we can
extract that there are three big contributors to the boot time, namely
\emph{Image download}, the \emph{Image duplication} and the \emph{Ephemeral
disk creation} steps.

In this first test, raw images were used, meaning that the duplication involved
the creation of a complete copy of the original image. This could be easily
diminished by using Copy on Write (CoW) images.

The support for CoW images is implemented in all of the most common hypervisors
(being the only difference the supported formats). Forcing the usage of CoW by
the Cloud middleware reduces considerably the overhead, since it is not needed
to duplicate the whole image container \cite{mcloughlin2008qcow2}. The
ephemeral disk (if it exists) can be also created using CoW, so its
contribution to the overhead will be diminished too. Therefore, one of the two
biggest contributors to the boot time for an instance can be easily shrink with
the adoption of CoW.

\begin{figure*}[ht!]
    \centering
    \includegraphics[width=1\linewidth]{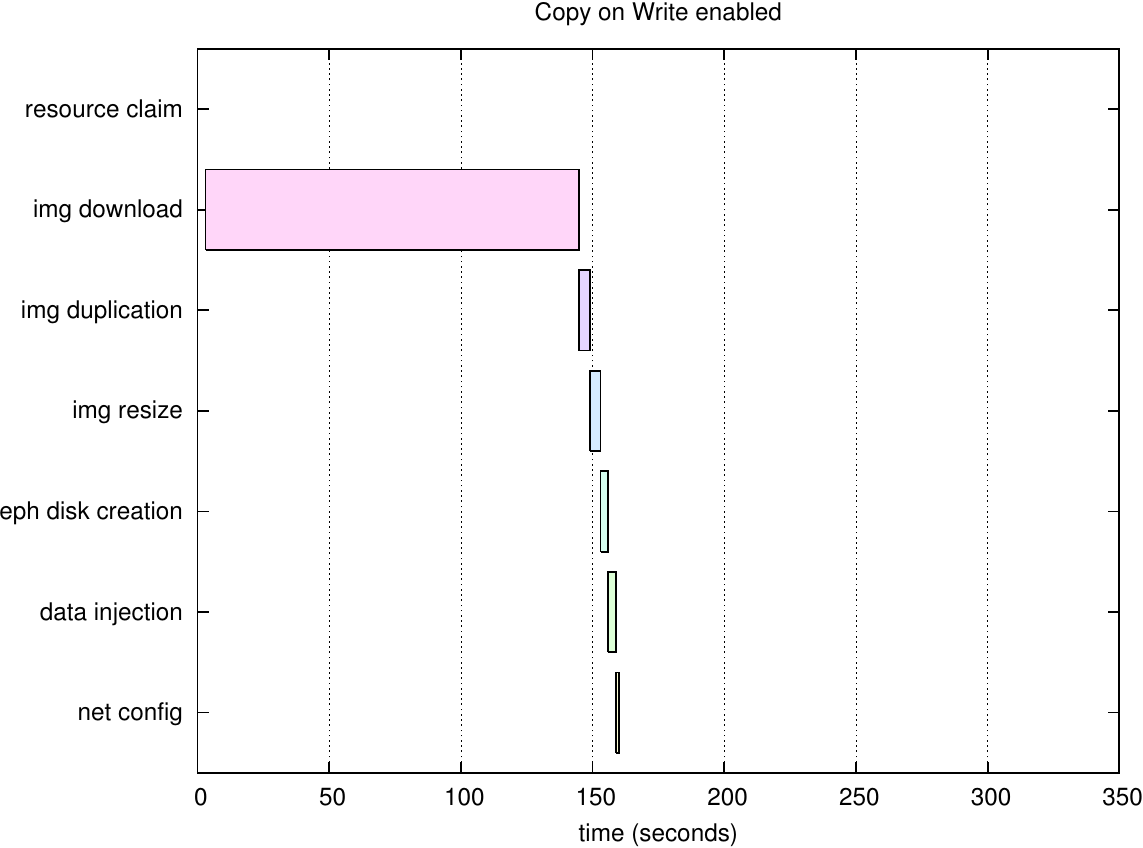}
    \caption{Chart of the boot process for one VM on an OpenStack cloud
        configured to use Copy on Write (CoW) images. The image used was 10GB
        large with an 80GB ephemeral disk.}
    \label{fig:bootchart-cow}
\end{figure*}

Figure~\ref{fig:bootchart-cow} shows the same request, when the cloud
infrastructure has been configured to use CoW images. As it is seen, two of the
three biggest penalties are reduced just by using Copy on Write images.

However, the \emph{Image download} still introduces the biggest penalty and,
unfortunately, this time is dependent on several factors:

\begin{itemize}
    \item The image delivery method used will have a large impact on the final
        time. It is not the same to download an image from a single central
        location that transfer it using peer-to-peer techniques.

    \item The amount of data being transferred and obviously the image size: if
        several hundred gigabytes need to be transferred over the network each
        time a machine is booted, the delay will be difficult to shrink.

    \item The size of the request. It is not the same to swap just a few
        virtual machines that spawning hundreds of VMs.

    \item The load on the implied systems: the network usage, catalog server
        and compute hosts load have an influence on the overall process.
\end{itemize}

Virtual machine images range from a few hundreds of Megabytes to several
Gigabytes \cite{Segal2010,Femminella2014}, hence an efficient image deliver
method should try to tackle as much factors as possible. It should try to use a
good image transfer method, should try to reduce the amount of data being
transferred and thus reduce the load on the system. It should be also able to
satisfy large requests, that are quite common on scientific workloads.  For
example, it is know that scientific communities often deploy a virtual cluster
to support their users \cite{Keahey2008,Afgan2010}, sets of machines to execute
a parallel application or workflow based applications \cite{Hardt2012}.

\section{Related work}
\label{sec:related}

Several authors have also identified the image deployment phase as the biggest
overhead to be solved when spawning VMs in a cloud infrastructure and have
proposed different solutions. In the following subsections we will describe
several of the proposals in the literature for addressing this issue.

One of the first approaches to reduce the image distribution time is to eliminate
the step itself. This could be accomplished by the usage of of a shared storage
(Section~\ref{sec:distribution:shared}) or by the pre-deployment or pre-fetch
of images (Section~\ref{sec:distribution:predeploy} and
Section~\ref{sec:distribution:prefetch} respectively). The election of a good
delivery method (Section~\ref{sec:distribution:ondemand}) is also crucial.
Finally, some authors point towards different and novel methods requiring
further developments (Section~\ref{sec:distribution:other}), that seem
promising.

\subsection{Shared storage}
\label{sec:distribution:shared}

This approach leverages the usage of a shared storage (such as access to a
Network Attached Storage (NAS) or a
Storage Area Network (SAN)) to eliminate at all image transfer. The
catalog and the nodes share the same storage backend, thus once an image is
uploaded to the system it will be directly available on the physical
hosts. This method may seem ideal, however, it still has some drawbacks:

\begin{itemize}
    \item The virtual machine disk is served over the network and nodes with an
        intensive Input/Output may underperform.

    \item It needs a dedicated and specialized storage system and network in
        order to not overload the instance's network with the access to the
        disks. This network needs to be properly scaled, meaning that a good
        performance access and acceptable reliability and availability are a
        must: if the shared storage does not perform as expected, it will
        become a bottleneck for the cloud infrastructure and will impact
        negatively on the virtual machines performance.

    \item If the system is not reliable or has a low availability, the images
        could not be accessed. Therefore, the IaaS resource provider needs
        to invest in having a good shared storage solution.

    \item The access to the shared storage by the physical machines (i.e. the
        hypervisor nodes) will consume resources and create undesirable VMM
        Noise. This VMM noise has been shown to have impact in the virtualized
        guests running on those hosts and is something to avoid in scientific
        computing environments
        \cite{Menon2005,Ferreira,Petrini:2003:CMS:1048935.1050204,gavrilovska2007high}.
\end{itemize}

\subsection{Image transfer improvements}

The use of shared storage may eliminate the image transfer into the nodes, but
as we explained it may not be desirable. In this Section we will discuss
several possibilities to decrease the image transfer time.

\subsubsection{On demand downloading}
\label{sec:distribution:ondemand}

If no shared storage is in place, the most common approach in many CMFs is to
transfer the images on-the-fly into the compute nodes when a request to launch
a specific machine is made.

As we already exposed in Section~\ref{sec:problem} the penalty introduced by
this method will vary according to the size of the image, the size of the
request, the network connectivity of the infrastructure, the load on the
catalog servers and the transfer protocol being used.

If the on demand download is the chosen option, the objective should be
reducing the image transfer time. In this line there is a clear trend towards
studying Peer-to-Peer (P2P) mechanisms in cloud infrastructures and
data-centers. Zhang Chen et al. \cite{ChenZ2009} proposed an effective approach for
virtual images provisioning based on BitTorrent. Laurikainen et al.  conducted
a research focused on the OpenNebula cloud middleware, taking only into account
the replacement of the native image transfer method by either BitTorrent or
Multicast \cite{Laurikainen2012}. Their conclusions showed that the existent
image transfer manager (based on SSH) was rather inefficient for large requests
and therefore it needed to be modified.

Wartel et al. studied BitTorrent among other solutions as the image transfer
method for their legacy CERN cloud infrastructure \cite{Wartel2010}. This study
shown a significant performance gain when using BitTorrent over the other
studied methods (that included multicasting). In the same line, Yang Chen et
al. have proposed a solution based on multicasting the images instead of a direct
download from the image catalog, in combination with a more efficient
scheduling \cite{Chen2009} algorithm. However, transfer an image using
multicast into the nodes implies that the server is initiating the transfer
(i.e. the server pushes the image into the nodes) instead of the image being
pulled from the hosts. This also forces that the deployment of the images is
synchronized, therefore introducing extra complexity to the scheduling
algorithms that must take this synchronization into account.

Once the image is downloaded into the node, this image can be cached and reused
afterwards in a subsequent request. This feature opens the door to the
pre-deployment of images and the image pre-fetch, that will be discussed in
Section~\ref{sec:distribution:predeploy} and
Section~\ref{sec:distribution:prefetch} respectively. Multicasting is an
interesting option for these two cases, since the deployment could be done in a
coordinated way, without interfering with the scheduling algorithms, but when
compared with multicast, using a P2P method introduces another advantage:
the nodes that have an image available are part of the P2P network,
participating actively in the transfer when a new request is made.

\subsubsection{Pre-deployment of images}
\label{sec:distribution:predeploy}

A different approach towards the elimination of the image transfer prior to the
image boot consists on the pre deployment of the whole or a portion of the
image catalog into the physical machines. In some environments this might be a
valid solution, but it is not affordable in large setups for several reasons.

First, in an infrastructure with a large catalog a considerable amount of disk
space will be wasted on the nodes. Considering that not all the images will be
spawned into all the nodes at a time, this resource consumption is not
affordable. Second, the pre-deployment process can overload the catalog
server when it is triggered if it is not properly scheduled or if the catalog
is too large.

A CMF using this method should also consider that a recently uploaded image may
not be immediately available to the user, since it has to be pre-seeded into
the nodes in advance, so an alternative, on-demand method should still be available.

\subsubsection{Smart pre-fetch}
\label{sec:distribution:prefetch}

Another possibility, related with the previous one, is performing a selective
pre-deployment of the images into the nodes (i.e. smart pre-fetch). Instead of
the passive deployment of the whole catalog (or a large portion of it) into the
nodes, the scheduler may chose to trigger a download of an image in advance,
so that it anticipates a user request.

\begin{figure}[ht!]
    \centering
    \includegraphics[width=1\linewidth]{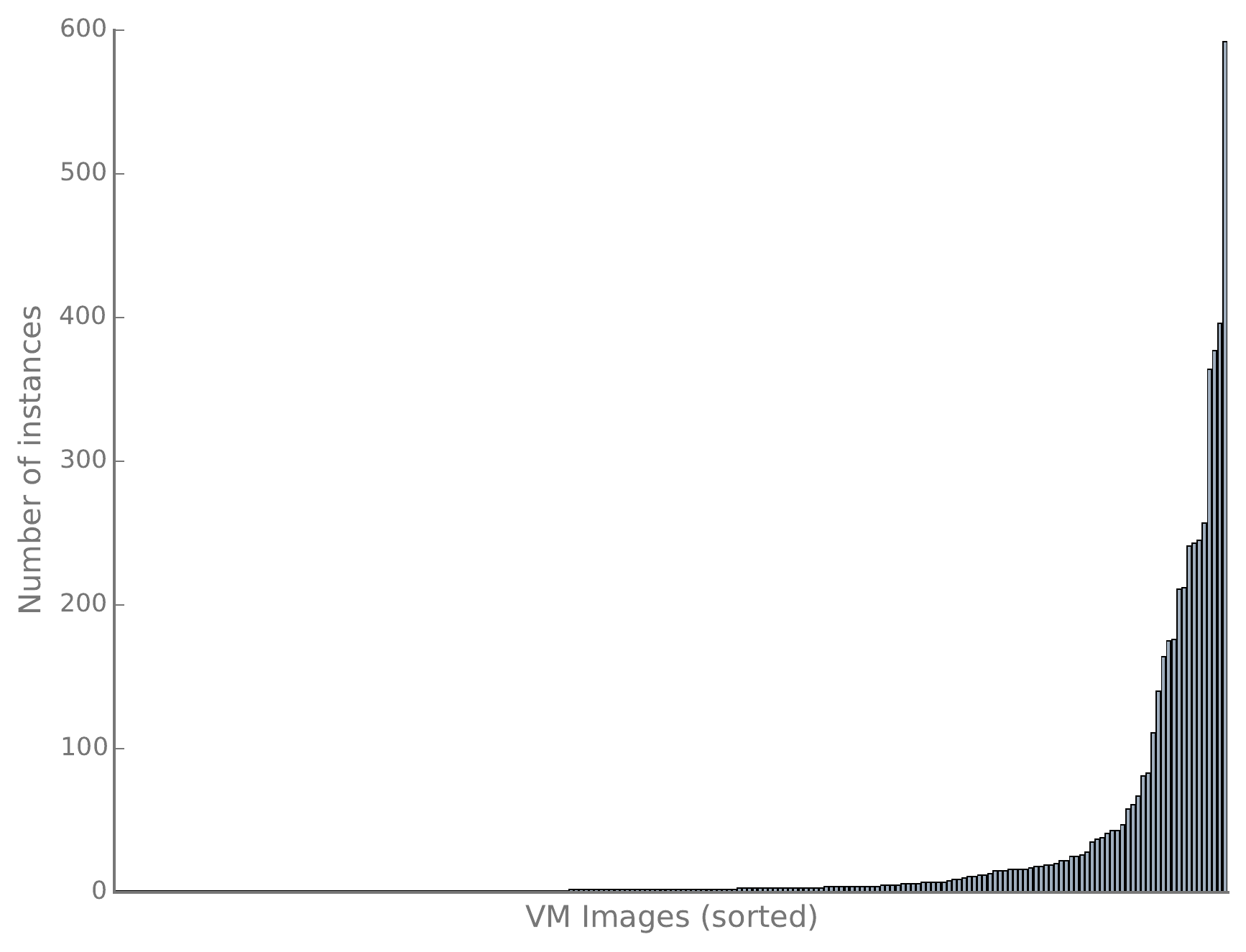}
    \caption{Image popularity based on the number of Virtual Machines spawned
        per image. Each bar represents a different image.}
    \label{fig:pop}
\end{figure}

Image popularity (i.e. how often an image is instantiated) can be used as
a parameter to decide which images to pre-fetch. A naive approach could be
summing up how many virtual machines have been instantiated from a given image.
Figure~\ref{fig:pop} shows the image \emph{popularity} for a set of $13500$
VMs execute by $150$ different users on a production cloud infrastructure
during one year in ascending order. The Y-axis shows the number of instances
that were based on a given image.

As it can be seen, even if this popularity calculation is too naive, a large
proportion of the spawned instances is spawned from a small number of images.
Some other authors have observed the same behaviour in some
related works, such as Peng et al. \cite{Peng2012}.
Therefore, if the CMFs could take advantage of the image popularity making
those VMs available on some nodes the efficiency of the image booting process
will improve.

\subsection{Other methods}
\label{sec:distribution:other}

Lagar-Cavilla et al. have developed Snowflock \cite{Lagar-Cavilla2011}, a new
model for cloud computing that introduces VM forking in a way similar to
the well known and familiar concept of process forking. This method permits the
cloning of
an already running VM into several identical copies. However this is not
transparent, and the users need to be aware of its semantics and program their
application accordingly.

Some other authors have chosen a totally different approach relying on the fact
that the image is not needed completely at once, therefore it can be divided
into smaller chunks that will be transferred when they are needed. Peng et al.
propose the usage of a collaborative network based on the sharing of similar
image chunks \cite{Peng2012}. In their studies, they found that this approach
was more efficient than the usage of a P2P network, but it requires a long
running preprocessing step. Moreover, this is true for the cases analyzed,
where the number of different VMs requested at a time was not big but this may
not apply to other cases, such as scientific cloud providers where the same
image may need to be spawned into several nodes.

The work from Nicolae et al. is also based on this approach.  They implemented
a self adaptive mechanism, based on lazy downloads of image chunks, based on
previously recorded access patterns \cite{Nicolae2011}.

\section{Transfer method evaluation}
\label{sec:evaluation}

There is no silver bullet for solving the image distribution problems, since
all of the presented schemes have their advantages and disadvantages. In some
situations, the usage of a shared backend may be the best solution but it would
not fit others. For example, sites deploying virtual machines that need high
availability may already use a shared backend so that it is possible to quickly
recover a running machine from a failure, whereas sites devoted to HTC and HPC
computing may not find this deployment appropriate. In this Section we evaluate
several image transfer methods in a Cloud Management Framework.

\subsection{Experimental setup}
\label{sec:testbed}

The tests were performed in a dedicated cloud testbed running only these
workloads.  It comprises a \emph{head node} hosting all the required services
to manage the cloud infrastructure, an \emph{image catalog} server and 24
\emph{compute nodes} that will eventually host the spawned virtual machines.
All of them are identical machines, with two 4-core Intel\circledR
Xeon\circledR E5345 2.33GHz processors, 16GB of RAM and one 140GB, 10.000 rpm
hard disk.

The network setup of the testbed consists on two 10GbE switches, interconnected
with a 10GbE link. All the hosts are evenly connected to these switches using a
1GbE connection.

The operating system being used for these tests is an Ubuntu Server 14.04 LTS,
running the Linux 3.8.0 Kernel. In order to implement the solution proposed we
have used the OpenStack \cite{web:openstack} cloud middleware, in its Icehouse
(2014.1) version.

In order to execute the same tests easily we used a benchmarking as a service
product developed for OpenStack: Rally \cite{web:rally}. This tool allows for
the definition and repetition of benchmarks, so that the benchmarking tests can
be reproduced later on.

OpenStack's default method for distributing the images into the nodes is an
on-demand deployment: the images are fetched from the catalog when the
new virtual machine is scheduled into a compute (physical) node and its image
cannot be found on that host.

The catalog service component (whose codename is Glance) stores the images
using one of the many available backends, but independently of the backend
used, the default transfer method is HTTP.  When Glance stores the images in a
filesystem it is possible to setup a shared filesystem so that the space where
the images are stored by glance are available on the compute nodes. Other
backends make possible to distribute the images over the network using
different protocols and methods (for example, using the Ceph Rados Block
Devices (RBD)). However, since we wanted to test the influence of the transfer
from the catalog to the nodes, the default method was used.

\subsection{Test results}

In order to evaluate the effect of the image transfer method we decided to
stress the system, making requests that involved fetching a large number
of images, as described in Table~\ref{tab:exp setup}, using several
methods: HTTP, FTP and BitTorrent. We used 5GB images and the
scheduler was configured to evenly distribute the images among the hosts in the
cluster in order to maximize the effect of the image transfer on the nodes. All
the tests were done by triplicate.

\begin{table}
    \centering
    \begin{tabular}{lccc}
        \toprule
        Name  & VMs per host & Different images & \# of VMs \\
        \midrule
        1x192 & 8 & 1  & 192 \\
        2x96  & 8 & 2  & 192 \\
        4x48  & 8 & 4  & 192 \\
        8x24  & 8 & 24 & 192 \\
        \bottomrule
    \end{tabular}
    \caption{Request characteristics.}
    \label{tab:exp setup}
\end{table}

\subsubsection{HTTP transfer}
\label{sec:http}

\begin{figure}[h!]
    \centering
    \includegraphics[width=1\linewidth]{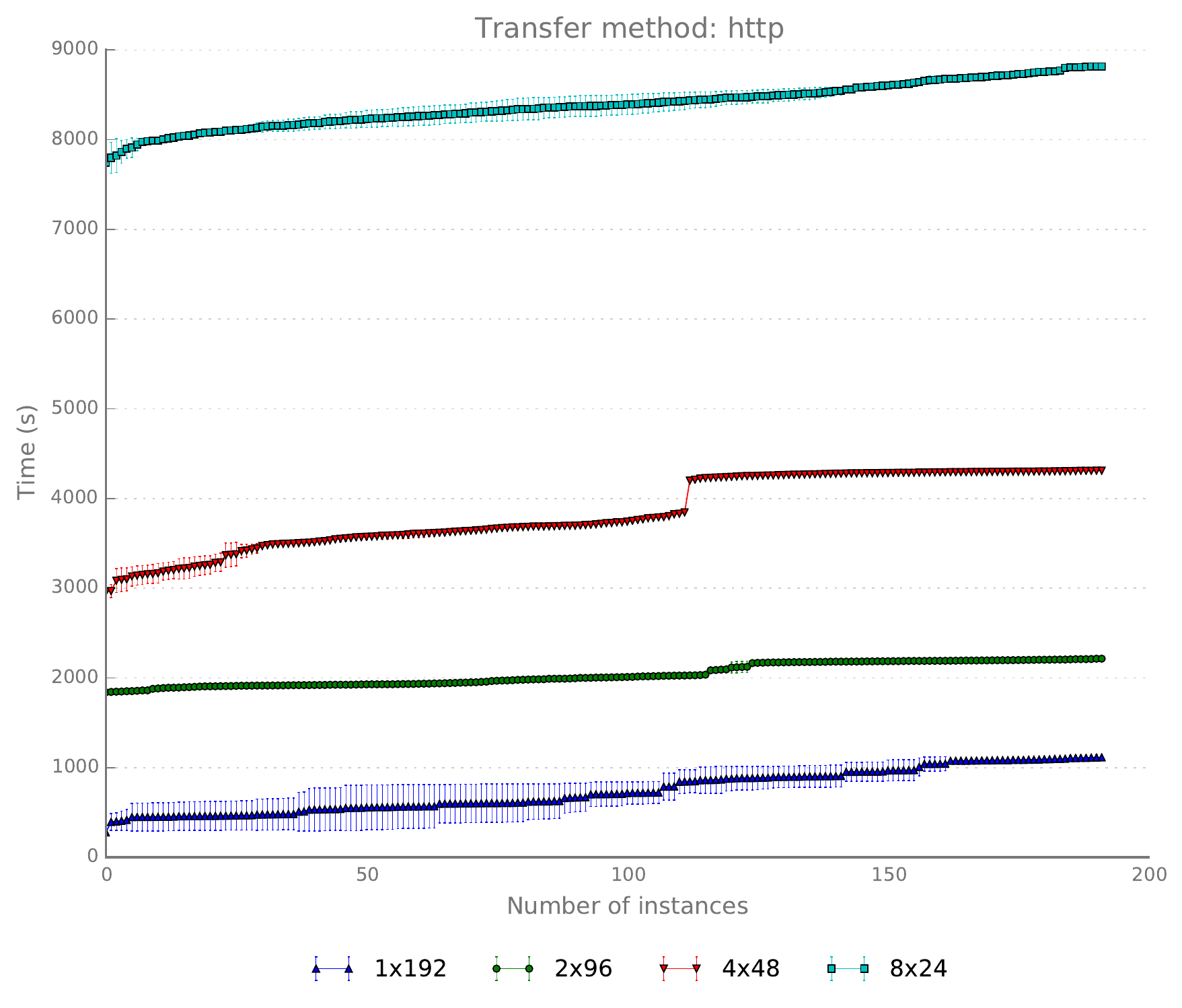}
    \caption[Waiting time in function of the number of instances requested]{
        Waiting time in function of the number of instances requested when the
        images are fetched using HTTP. 1x192
        means 1 request of 192 machines using the same image; 2x96, 2 requests
        of 96 machines using two different images, 4x48, 4 requests of 48
        machines with four different images; and 8x24 8 requests of 24 machines
        with eight different images. }
    \label{fig:http requests}
\end{figure}

In the first place we transferred the images using HTTP, since it is the default
image transfer method available on OpenStack.
Figure~\ref{fig:http requests}, shows the required time to boot the virtual
machines for each of the requests in Table~\ref{tab:exp setup}.

The best scenario in these tests is where a user requests a single image (1x192
in Figure~\ref{fig:http requests}). This is mainly because of the effect of the
cache that is available in each of the nodes. Once the image is downloaded in a
node, all the subsequent virtual machines can be spawned using that cached
image (this fact is also true for the other studied methods). The worst scenario is when the user requested 8 groups if 24 virtual
machines (8x24 in Figure~\ref{fig:http requests}), since all the 8 images had
to be downloaded into each of the nodes.

\subsubsection{FTP transfer}
\label{sec:ftp}

\begin{figure}[h!]
    \centering
    \includegraphics[width=1\linewidth]{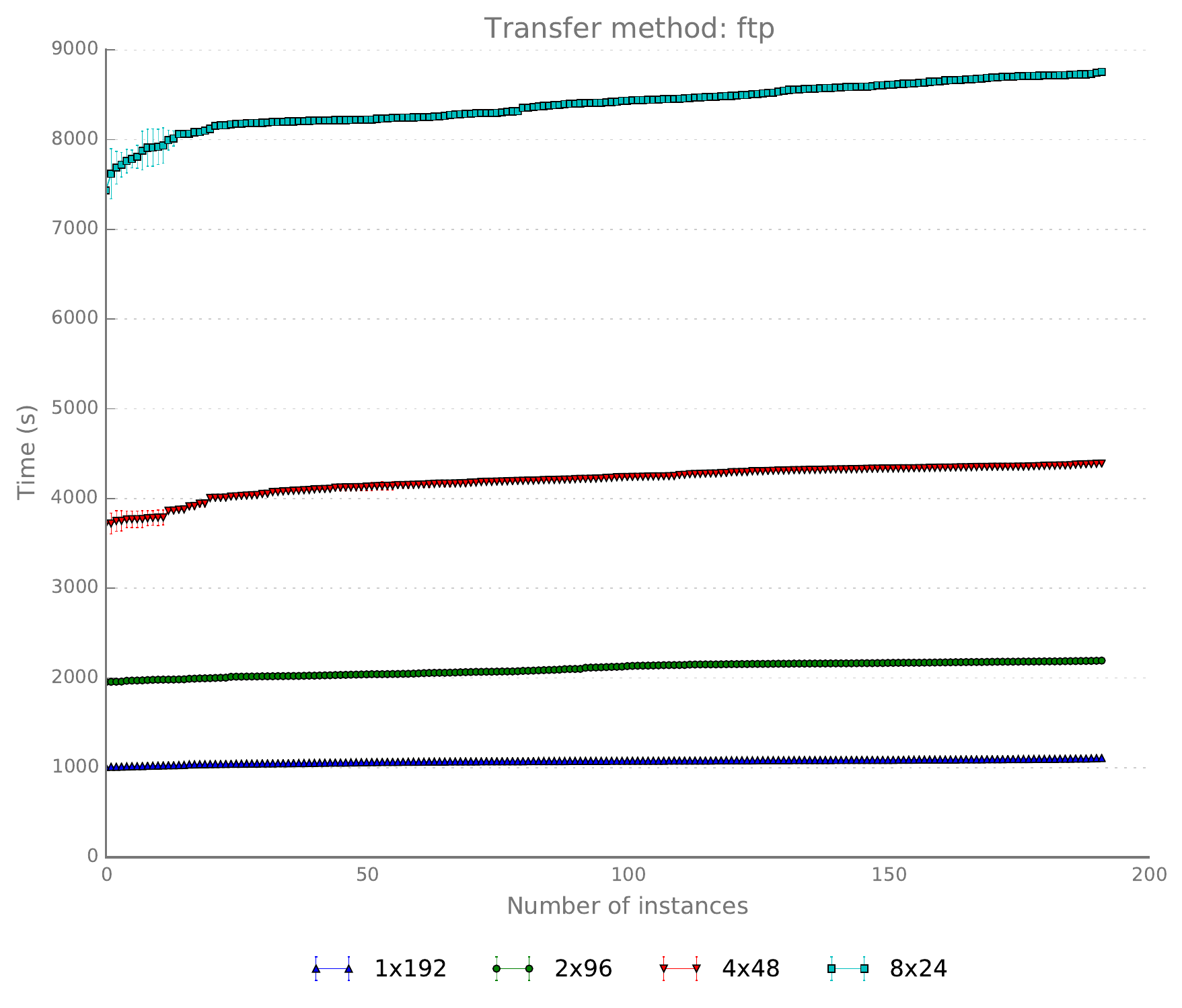}
    \caption[Waiting time in function of the number of instances requested]{
        Waiting time in function of the number of instances requested when the
        images are fetched using FTP. 1x192
        means 1 request of 192 machines using the same image; 2x96, 2 requests
        of 96 machines using two different images, 4x48, 4 requests of 48
        machines with four different images; and 8x24 8 requests of 24 machines
        with eight different images.}
    \label{fig:ftp requests}
\end{figure}

As a second step we decided to substitute the built-in HTTP server with a
dedicated FTP server, and use the File Transfer Protocol (FTP) instead.
Figure~\ref{fig:ftp requests} shows again the results for the requests in
Table~\ref{tab:exp setup}.

As it can be seen, the boot time is almost the same for both methods, being FTP
more homogeneous over HTTP, resulting in a most uniform boot time for the
machines.

\subsubsection{BitTorrent deployment}
\label{sec:bt}

Both the HTTP (Section~\ref{sec:http}) and FTP (Section~\ref{sec:ftp}) are based on a centralized
client-server model. In order to see how the system performs using a
peer-to-peer (P2P) model we adapted OpenStack image delivery method to use
BitTorrent. We chose it for several reasons: it is a protocol designed for to reduce the
impact of transferring large amounts of data over the network
\cite{cohen2008bittorrent}; it is widely used in a
daily basis and there is a wide range of libraries, clients and applications
available; moreover, due to this lively implementation ecosystem, we found
that it could be easily integrated into OpenStack.

We chose libtorrent \cite{web:libtorrent} as the implementation for our tests.
libtorrent has Python bindings, and since OpenStack is written entirely in
Python it was easily integrable. Our \emph{swarm} used the BitTorrent Distributed
Hash Table (DHT) extension, so that we could use tracker-less torrents, although
it is perfectly feasible to run a tracker. We configured the clients to run
only 3 concurrent active downloads, since in preliminary tests we observed this
was the best choice for our infrastructure.

\begin{figure}[h!]
    \centering
    \includegraphics[width=1\linewidth]{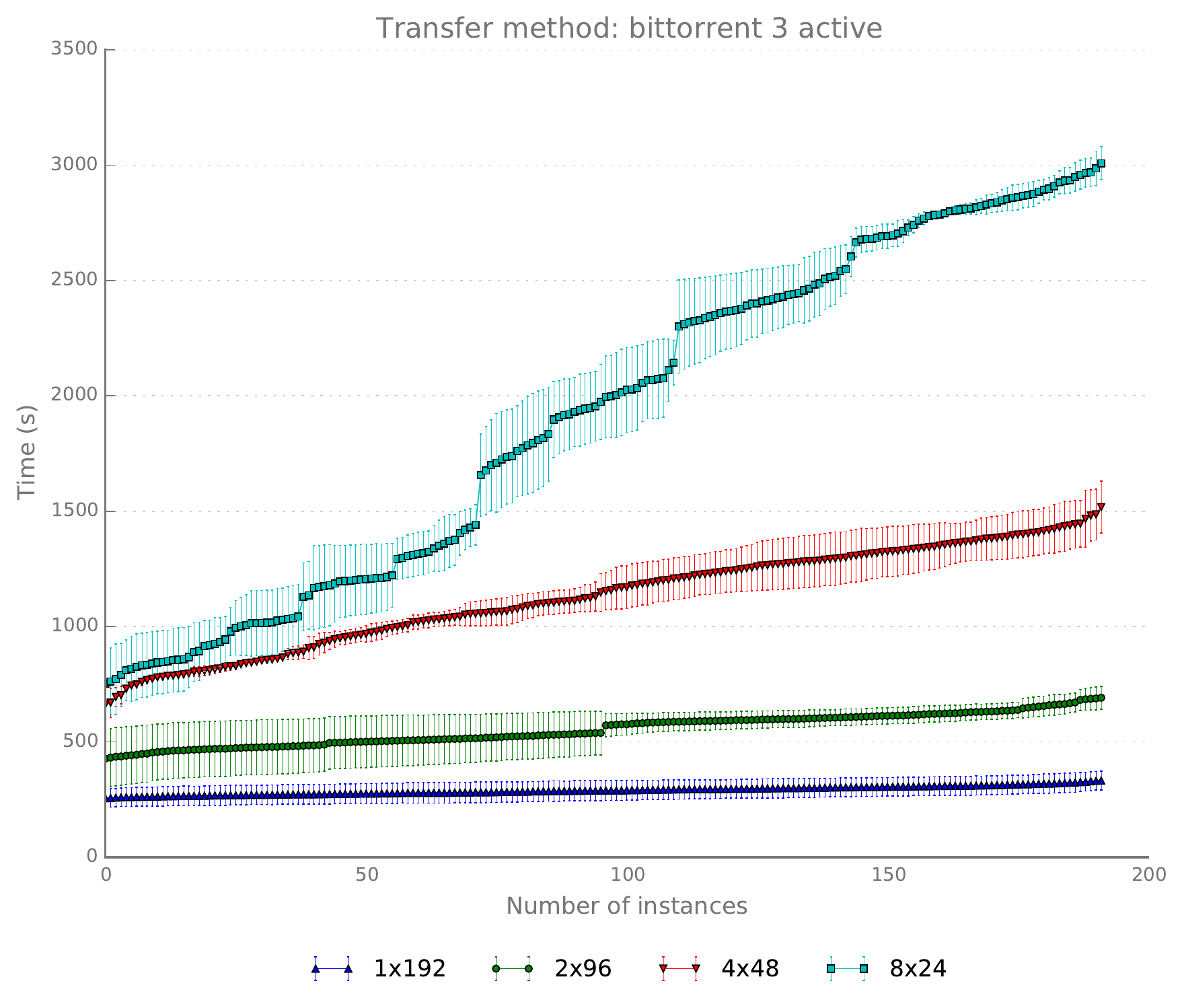}
    \caption[Waiting time in function of the number of instances requested]{
        Waiting time in function of the number of instances requested when the
        images are fetched using BitTorrent. 1x192
        means 1 request of 192 machines using the same image; 2x96, 2 requests
        of 96 machines using two different images, 4x48, 4 requests of 48
        machines with four different images; and 8x24 8 requests of 24 machines
        with eight different images.}
    \label{fig:bt requests}
\end{figure}

The results for serving the same requests as in the HTTP and FTP cases
are show in Figure~\ref{fig:bt requests}.

In our implementation a new torrent is generated whenever a new image is
uploaded to the catalog. The torrent metadata is stored along with the ordinary
image metadata so that whenever a download of this image is requested, both the
normal HTTP and the torrent's magnet link are provided to the compute node. If
the node needs to download the image, and a magnet link is available, this
\emph{peer} (i.e. a BitTorrent client) will join the \emph{swarm} (i.e. all
peers sharing a torrent). Due to the segmented file transfer that BitTorrent
implements, this \emph{peer} is able to \emph{seed} (i.e. send its available
data) the received data to the other peers. This way, the original seeder of
the image (i.e. the catalog server) is freed from sending that portion to every
peer of the network.

\subsection{Result comparison}
\label{sec:results}

\begin{figure}[h!]
    \centering
    \includegraphics[width=1\linewidth]{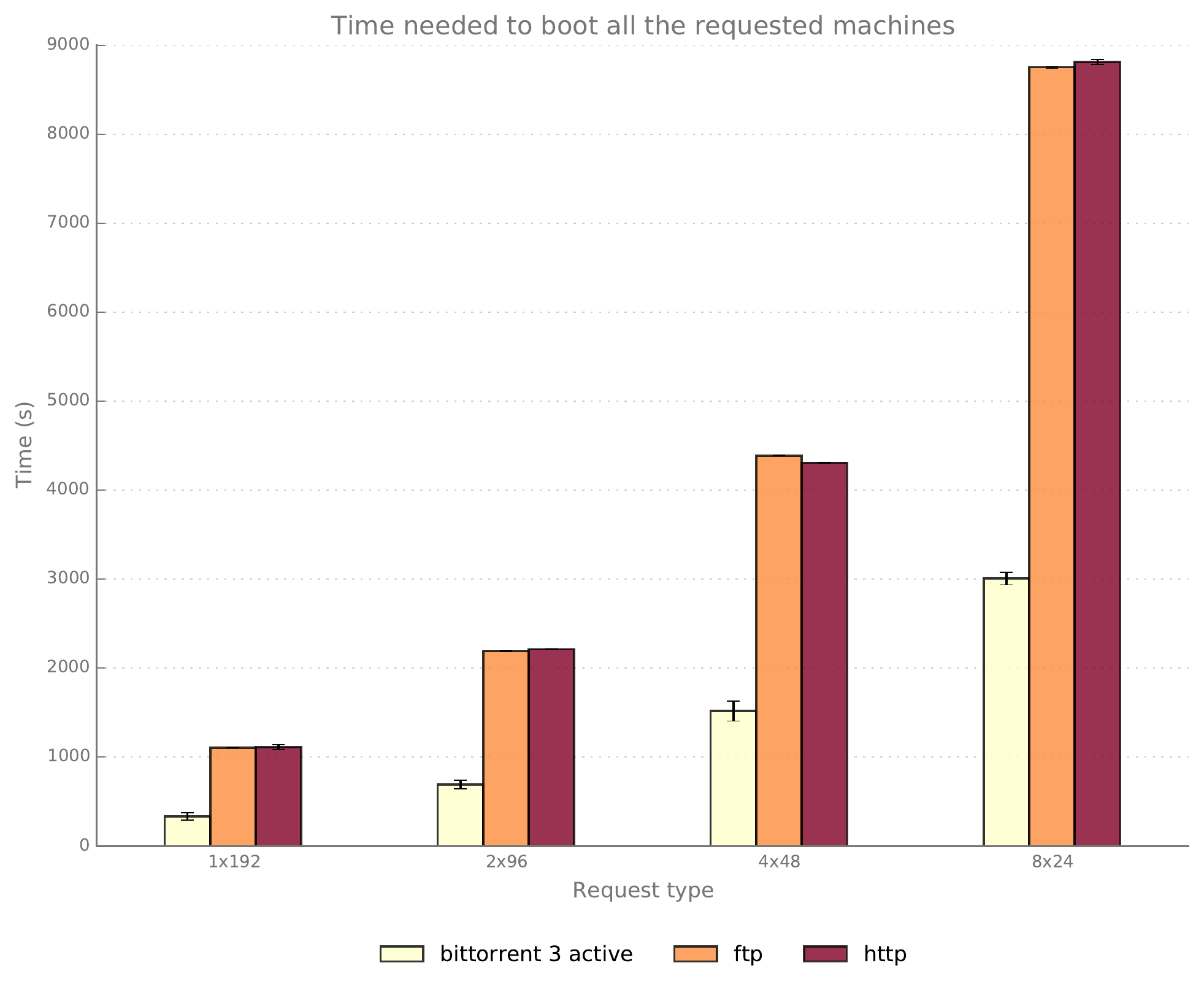}
    \caption{Seconds elapsed from request until all the machines were
    available.  The VMs were based on a 5GB image, and they were spawned on 24
    hosts.}
    \label{fig:start_time_multiple}
\end{figure}

A comparison of the three methods evaluated (that is, transfer the images using
HTTP, FTP and BitTorrent, and profit from the images caching) is shown in
Figure~\ref{fig:start_time_multiple}.

Both FTP and HTTP threw similar results, being those limited by the bandwidth
of the server node. Using BitTorrent, there is a significant transfer time
reduction.  In the worst scenario (8x24: running 192 virtual machines,
distributed in 8 different images in 24 nodes) it was possible to start the 192
machines at approximately one third of the time required to run those machines
using HTTP or FTP.

\begin{figure}[h!]
    \centering
    \includegraphics[width=1\linewidth]{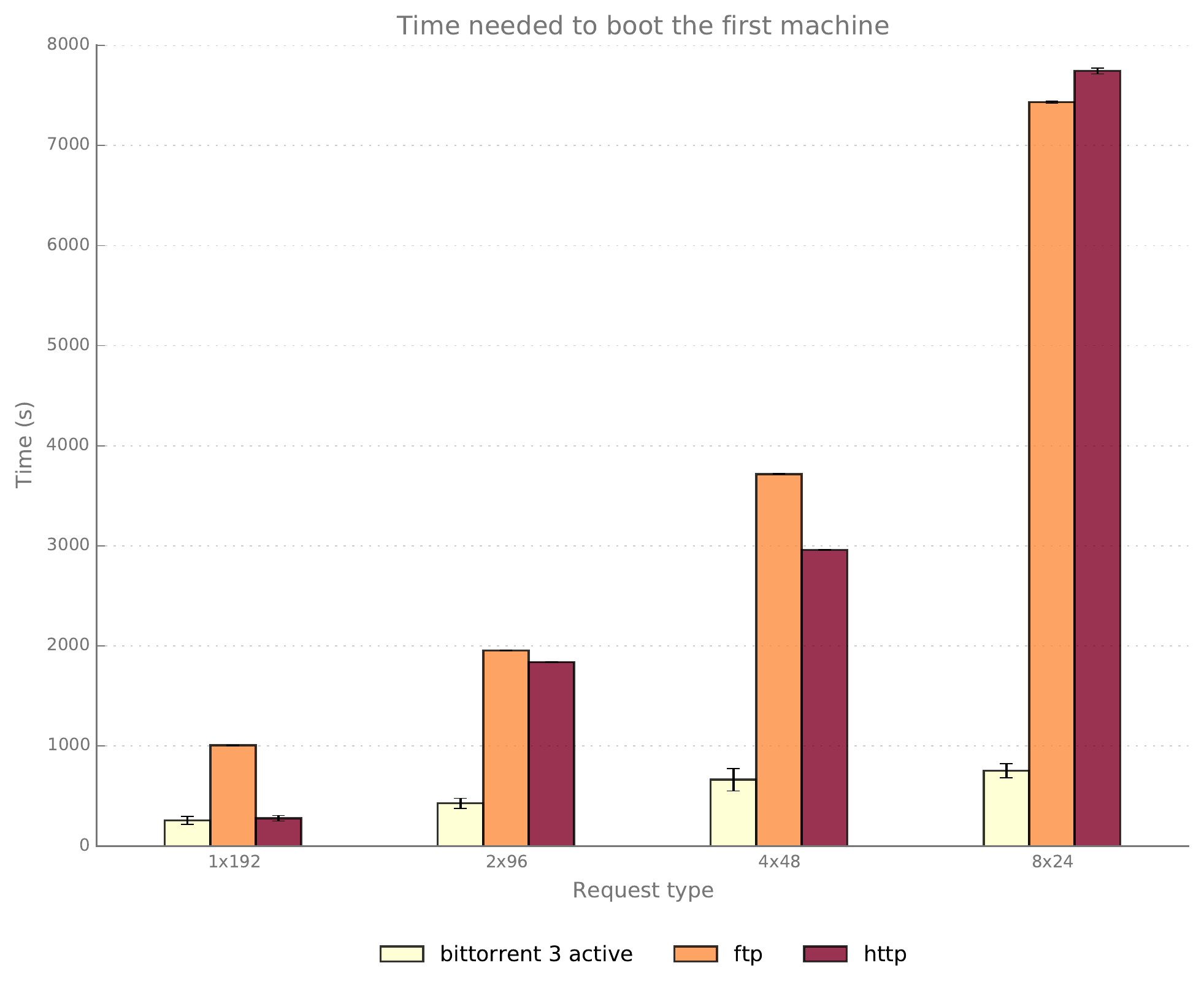}
    \caption{Seconds elapsed from request until the first machine of the
        request is available. The VMs were based on a 5GB image, and they were
        spawned on 24 hosts.}
    \label{fig:start_time_multiple_first}
\end{figure}

If we take into account the boot time for the first machine of the request we
can find interesting results. Figure~\ref{fig:start_time_multiple_first}
shows the elapsed time until the first machine is available. In
this case, BitTorrent also outperforms the other transfer methods, making
possible to deliver the machines earlier to the users except in the case of
transferring only one image into all the nodes. In this case, HTTP and
BitTorrent throw similar results.

Another important fact is that the adoption of BitTorrent not only has the
effect of reducing the transfer time, but it also reduces the load of
the catalog server. Since the image distribution leverages the advantages of
the P2P network, where all the nodes participate in the transfer, the catalog
does not need to transfer all the data to all of the nodes.

\begin{figure}[ht!]
    \centering
    \begin{subfigure}{0.7\textwidth}
        \centering
        \includegraphics[width=\linewidth]{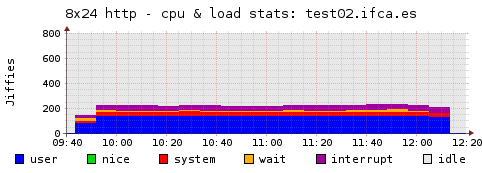}
        \caption{HTTP.}
        \label{fig:cpu:http}
    \end{subfigure}
    \\
    \begin{subfigure}{0.7\textwidth}
        \centering
        \includegraphics[width=\linewidth]{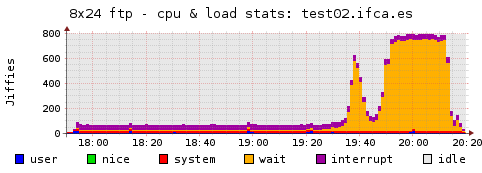}
        \caption{FTP.}
        \label{fig:cpu:ftp}
    \end{subfigure}
    \\
    \begin{subfigure}{0.7\textwidth}
        \centering
        \includegraphics[width=\linewidth]{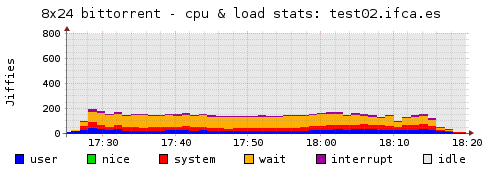}
        \caption{BitTorrent.}
        \label{fig:cpu:bt}
    \end{subfigure}
    \caption{CPU usage for a 192 VMs request using 8 different images (8x24).}
    \label{fig:cpu}
\end{figure}

\begin{figure}[ht!]
    \centering
    \begin{subfigure}{0.7\textwidth}
        \centering
        \includegraphics[width=\linewidth]{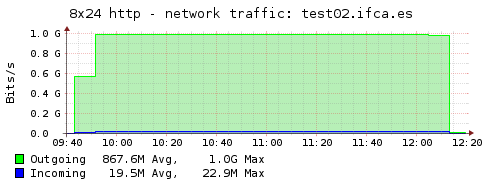}
        \caption{HTTP.}
        \label{fig:net:http}
    \end{subfigure}
    \\
    \begin{subfigure}{0.7\textwidth}
        \centering
        \includegraphics[width=\linewidth]{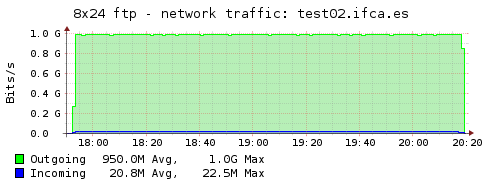}
        \caption{FTP.}
        \label{fig:net:ftp}
    \end{subfigure}
    \\
    \begin{subfigure}{0.7\textwidth}
        \centering
        \includegraphics[width=\linewidth]{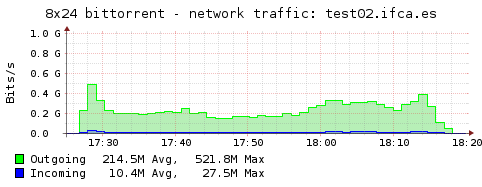}
        \caption{BitTorrent.}
        \label{fig:net:bt}
    \end{subfigure}
    \caption{Network usage for a 192 VMs request using 8 different images (8x24).}
    \label{fig:net}
\end{figure}

As it can be seen in Figure~\ref{fig:cpu} and Figure~\ref{fig:net}, using
BitTorrent makes possible to satisfy the same request at a fraction of the CPU
usage and specially network bandwidth when compared with HTTP and FTP,
resulting in a better utilization of the resources.

However, using BitTorrent has its drawbacks also. It needs another running
service (a tracker, although it could be avoided using a Distributed Hash Table
(DHT)). Moreover, the creation of a torrent file whenever a new machine image is
added to the catalog takes a considerable amount of time and resources, growing
with the size of the file. Therefore the torrent will not be available as soon
as the image is uploaded, but a lapse of time will be introduced. Since this
operation is done only once in the lifetime of a virtual machine it can be
considered as part of the initial upload process.

\section{Efficient image distribution}
\label{sec:cache scheduler}

In the previous section we have made emphasis in the effect of the image
distribution method on the boot time for a virtual machine. In all of the
presented these tests we have started from a clean environment, meaning that
there were no images cached in the nodes. The tests were designed to stress the
infrastructure so that the image transfer effects could be clearly noticed. In
this section we will evaluate the effect of taking into account the images
cached in a physical node when making scheduling decisions under more realistic
scenarios.

The default scheduling process in OpenStack has two steps: filtering and weighting.

The first step is the filtering phase. The scheduler applies a concatenation of
filter functions to the initial set of available hosts, based on the host
properties. When the filtering process has concluded, all the hosts in the
final set are able to satisfy the user request. At this point, the weighting
process starts so that the best suited host is selected.

The scheduler will apply to each of the hosts the same set of weighers
functions $\mathrm{w}_i(h)$ for each host $h$. Each of those weigher functions
will return a value considering the characteristics of the host received as
input parameter. Therefore, total weight $\Omega$ for a node $h$ is calculated
as follows:

$$
\Omega = \sum^n{m_i\cdot \mathrm{N}{(\mathrm{w}_i(h))}}
$$

Where $m_i$ is the multiplier for a weighter function,
$\mathrm{N}{(\mathrm{w}_i(h))}$ is the
normalized weight between $[0, 1]$ calculated via a rescaling like:

$$
\mathrm{N}{(\mathrm{w}_i(h))} = \frac{\mathrm{w}_i(h)-\min{W}}{\max{W} - \min{W}}
$$

where $\mathrm{w}_i(h)$ is weight function, and $\min{W}$,
$\max{W}$ are the minimum and maximum values that the weigher has assigned for
the set of weighted hosts.

Once the set of hosts have weights assigned to them, the scheduler will select
the host with the maximum weight and will schedule the request into it. Eventually,
if several nodes have the same winner weight, the final host will be randomly
selected from that set.

In order to evaluate how the cache could improve the boot time, we tested
four different scenarios: using the OpenStack's default scheduling algorithm
and using a cache-aware scheduler; using both HTTP and BitTorrent as the
transfer methods. This we we could asses not only the effect of the cache but
also the transfer method.

In our test environment all the hosts have the same hardware characteristics,
so when they are empty they are equally eligible for running a machine. As
explained, the nodes will get the same weight and finally a random selection is
done. Therefore it is possible that a machine is scheduled in a node that does
not have the image available, when there is another node with the same weight
with the image cached. In the best case, the image is transferred only once
(that, is for the first request), whereas in the worst case the image will have
to be transferred every time it is used.

By default OpenStack has an image cache in each of the nodes, but the scheduler
does not take it into account when selecting the host that will execute a
machine. We developed several modules for OpenStack, allowing to weight the
hosts taking into account their cached images. First of all, the nodes have
report their cached images back to the scheduler. Afterwards, the cache
weigher will simply weight the nodes as follows:

$$
\mathrm{w}_{\mathrm{cache}}(h) = \left\{
    \begin{array}{ll}
        1 \quad \text{if image is cached} \\
        0 \quad \text{otherwise}
    \end{array} \right.
$$

We did not apply any other sanity check in the weigher since this is not the
purpose of our function (there are specific weighers and filters that should
prevent to overload a host).

Therefore, with the above configuration in the cache-aware tests, the images
were only transferred the first time they are scheduled, since all the
subsequent requests will be scheduled in any of those hosts.

\subsection{Evaluation}

In order to make a realistic evaluation, we executed different simulated
request traces for each of the scenarios described before: that is, an
scheduler with and without cache, using HTTP and BitTorrent.

We generated two arrival patterns using an exponential distribution
\cite{knuth1981art}: one for a rate of 80 machines per hour and a second one
for 100 machines per hour. For each of the requests we assigned an image chosen
randomly from a given set of 4 images. Finally, the two resulting traces were
executed in each of the four scenarios.

\begin{figure}[h!]
    \centering
    \includegraphics[width=1\linewidth]{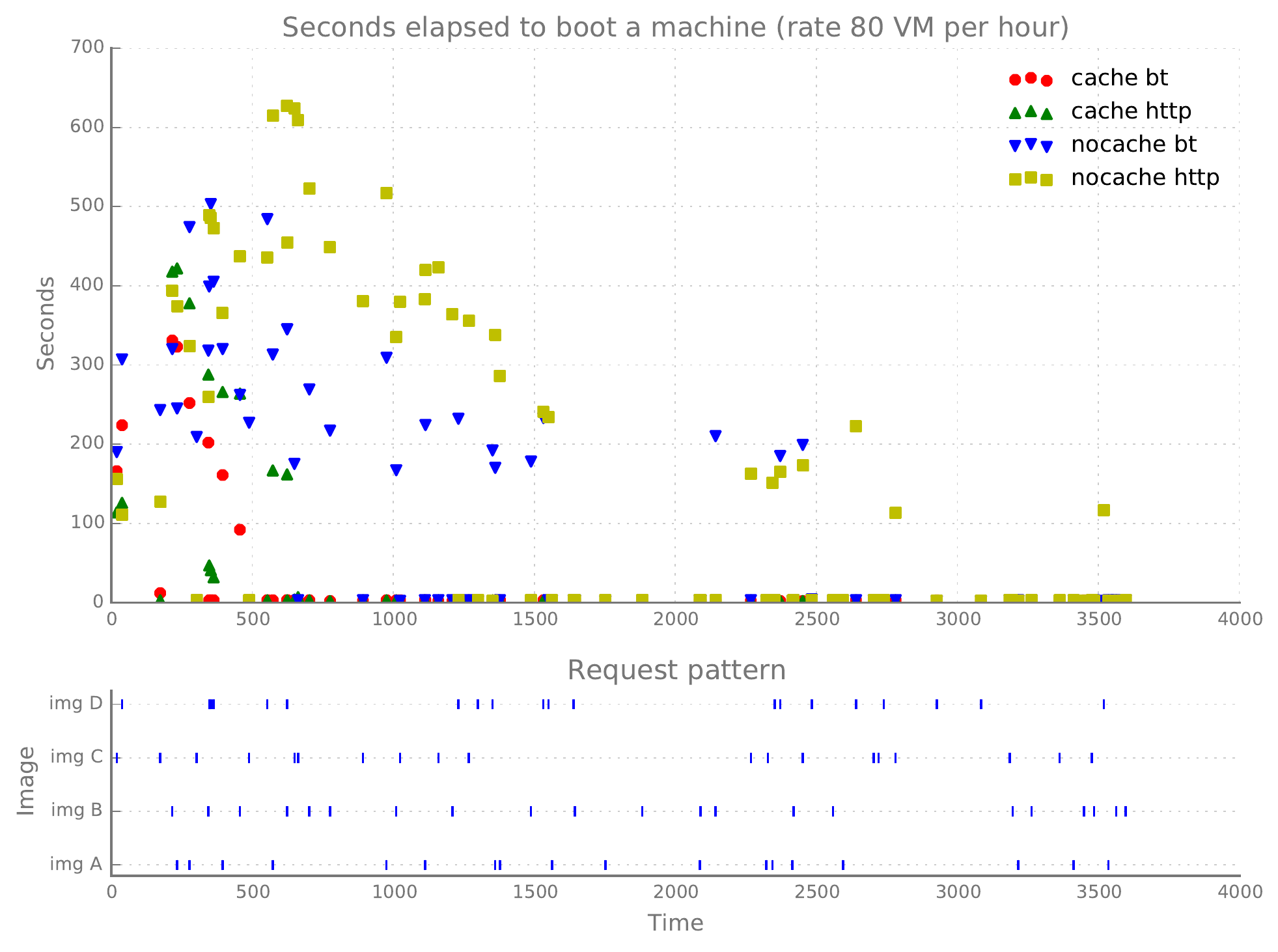}
    \caption{Seconds elapsed to boot a machine for 80 requests during 1 hour,
        with the corresponding requests trace.
        \texttt{nocache http} and
        \texttt{nocache bt} refer to the default scheduling method using HTTP
        and BitTorrent respectively, whereas \texttt{cache http} and
        \texttt{cache bt} refer to the cache-aware scheduler, using HTTP and
        BitTorrent respectively.
}
    \label{fig:scatter 80}
\end{figure}

\begin{figure}[h!]
    \centering
    \includegraphics[width=1\linewidth]{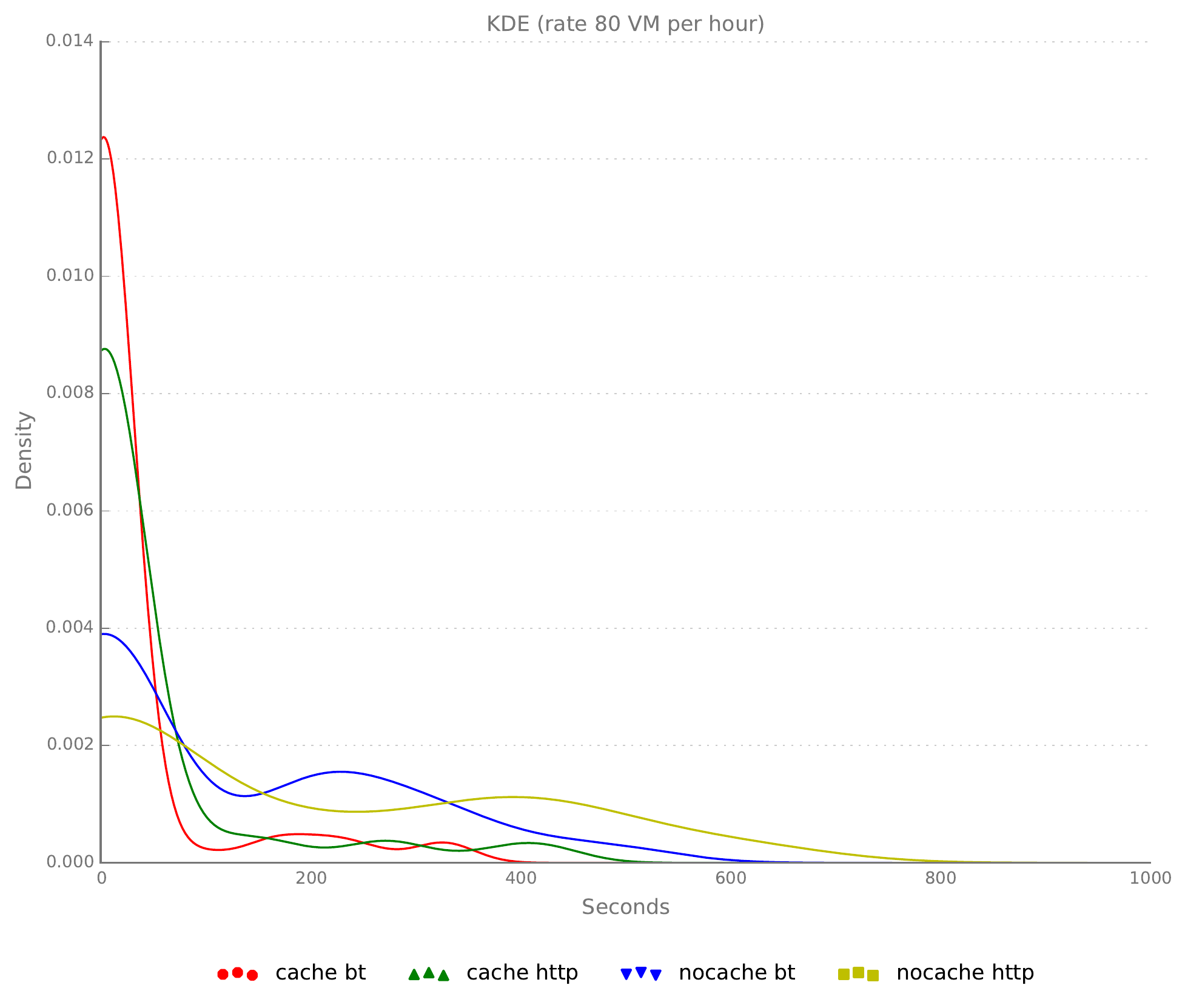}
    \caption{Kernel density estimation for the time elapsed to boot the
        requests in Figure~\ref{fig:scatter 80}.
        \texttt{nocache http} and
        \texttt{nocache bt} refer to the default scheduling method using HTTP
        and BitTorrent respectively, whereas \texttt{cache http} and
        \texttt{cache bt} refer to the cache-aware scheduler, using HTTP and
        BitTorrent respectively.
}
    \label{fig:density 80}
\end{figure}

Figure~\ref{fig:scatter 80} shows the scatter plot of the seconds needed to
boot each of the requests and its respective request pattern for 80 machines at
an arrival rate of 80 machines per hour. Figure~\ref{fig:density 80} shows the
kernel density estimation of the test.

\begin{figure}[h!]
    \centering
    \includegraphics[width=1\linewidth]{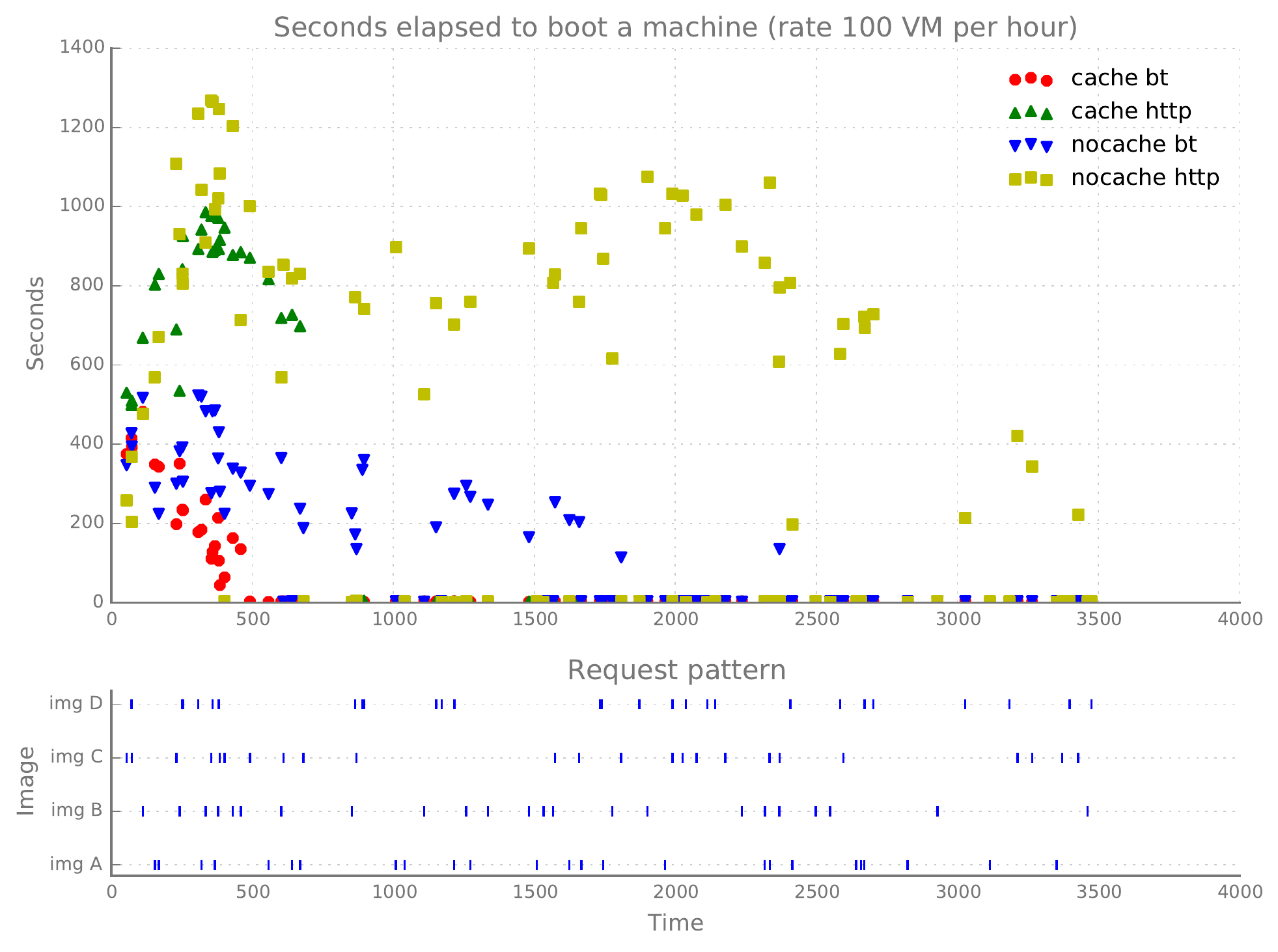}
    \caption{Seconds elapsed to boot a machine for 100 requests during 1 hour,
        with the corresponding requests trace.
        \texttt{nocache http} and
        \texttt{nocache bt} refer to the default scheduling method using HTTP
        and BitTorrent respectively, whereas \texttt{cache http} and
        \texttt{cache bt} refer to the cache-aware scheduler, using HTTP and
        BitTorrent respectively.
}
    \label{fig:scatter 100}
\end{figure}

\begin{figure}[h!]
    \centering
    \includegraphics[width=1\linewidth]{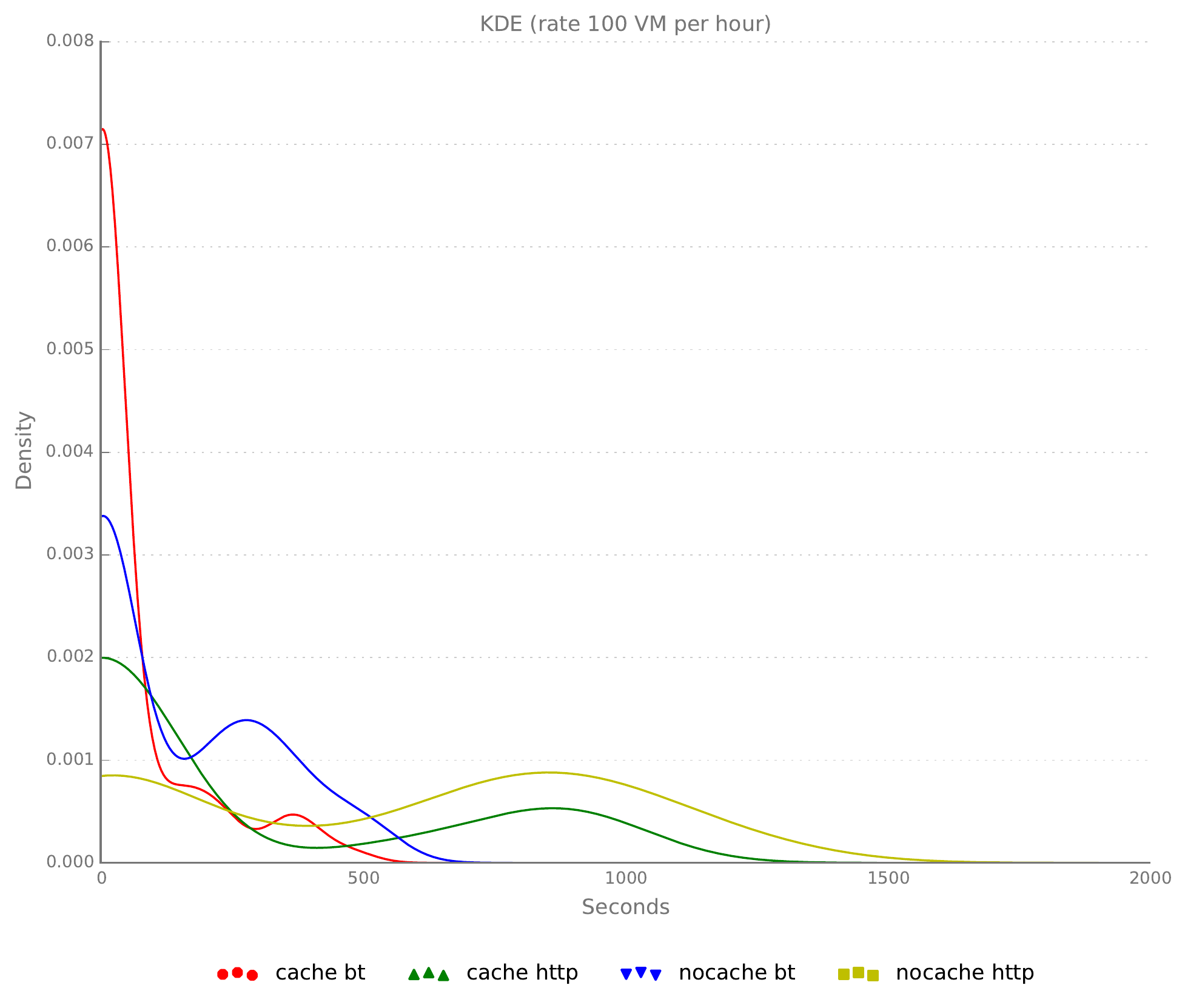}
    \caption{Kernel density estimation for the time elapsed to boot the
        requests in Figure~\ref{fig:scatter 100}.
        \texttt{nocache http} and
        \texttt{nocache bt} refer to the default scheduling method using HTTP
        and BitTorrent respectively, whereas \texttt{cache http} and
        \texttt{cache bt} refer to the cache-aware scheduler, using HTTP and
        BitTorrent respectively.
}
    \label{fig:density 100}
\end{figure}

Besides, Figure~\ref{fig:scatter 100} contains the plot for 100 machines at an arrival
rate of 100 machines per hour, with the corresponding density function shown in
Figure~\ref{fig:density 100}.

As it can be seen in both Figures \ref{fig:scatter 80} and
\ref{fig:scatter 100}, in all evaluated scenarios the minimum values are similar
and very low due to the effect of the cache. In the cases when the scheduler
did not have this feature available there is still a random chance that a
machine is scheduled in a node with the image cached, thus the observed
results. The probability of using a node with the image already available
increases with time (more nodes have been used and therefore more nodes have
the image cached) and as a consequence the boot times for the last images was
lower. When the cache-aware scheduler was used, only the first machines started
require transfer to the nodes, hence the boot times are reduced to the minimum
early in the execution of the trace.

On the other hand, Figures \ref{fig:density 80} and \ref{fig:density 100}
thrown interesting results, considering the size of the requests. The best
results are always obtained when using BitTorrent and a cache-aware scheduler.
However, the next best case depends on the request pattern. In the case of a
rate request of 100 machines per hour, using BitTorrent without a cache is
better than using HTTP with a cache, but in the case of a rate of 80 machines
per hour it is better to use the later. This observation is due to the fact
that in the 100 machines case there is a large initial portion of images that
need to be transmitted if compared with the 80 machines case, (as depicted by
the dots between time $0$ and $500$ in Figures \ref{fig:scatter 80} and
\ref{fig:scatter 100}. Therefore BitTorrent outperforms HTTP, as already
explained in Section~\ref{sec:results}. The cache does not consider the images
that are being fetched, therefore the scheduler cannot take them into account.
As the 100 machines case requests machines at a higher rate they are being
scheduler when the images are not yet available, thus the observed results.

\subsection{Image pre-fetch}

\begin{figure}[h!]
    \centering
    \includegraphics[width=1\linewidth]{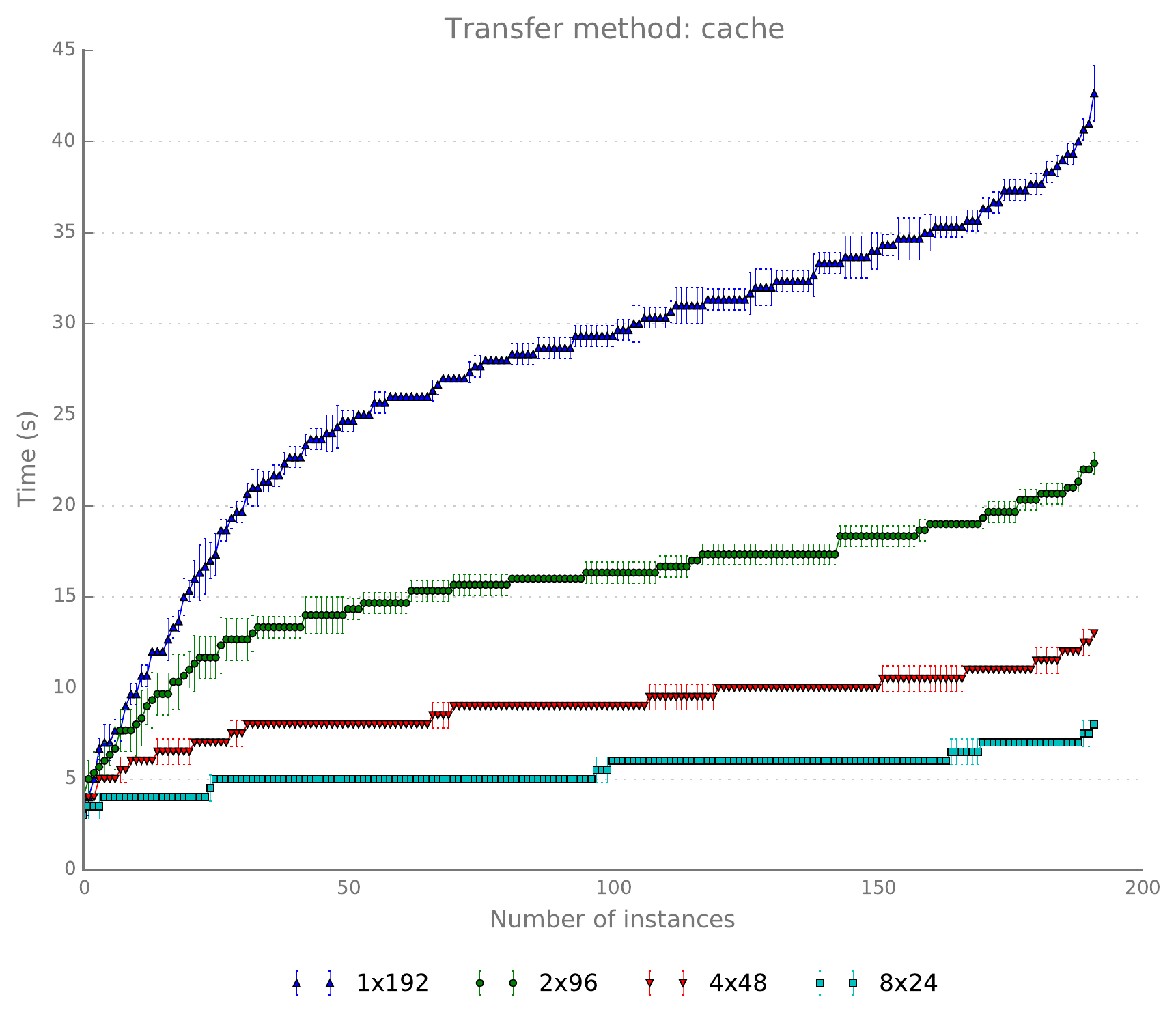}
    \caption[Waiting time in function of the number of instances requested]{
        Waiting time in function of the number of instances requested when the
        images are cached in the nodes. 1x192
        means 1 request of 192 machines using the same image; 2x96, 2 requests
        of 96 machines using two different images, 4x48, 4 requests of 48
        machines with four different images; and 8x24 8 requests of 24 machines
        with eight different images. }
    \label{fig:cache requests}
\end{figure}

As already explained, the usage of the cache with BitTorrent outperforms all of
the other methods. In order to evaluate its effect regarding the tests shown in
Section~\ref{sec:evaluation} we recreated the same requests from
Table~\ref{tab:exp setup} with the images already cached on the nodes.
Obviously, in this test we do not evaluate the penalty introduced by the image
transfer since there is no transfer at all, but it is interesting in order to
evaluate the overall performance of the system.  As it can be seen in
Figure~\ref{fig:cache requests}, the booting time was dramatically reduced in
all cases: booting all the 192 machines was done in less than 45 seconds as the
only delays introduced where due to the scheduling algorithm and the different
management operations.

\section{Conclusions and future work}
\label{sec:conclusions}

In this paper, we have evaluated several methods for the distribution of
virtual machine images into the compute nodes of a cloud infrastructure.
Although the work was performed using the OpenStack cloud middleware, the
results can be extrapolated to other CMFs using similar transfer methods.

Our experiments showed that composing a P2P network based on a well established
protocol such as BitTorrent is a simple, feasible and realistic solution to
decrease the burden on the server and to reduce the transfer time to a smaller
fraction of time.

Moreover, we have also evaluated the usage of an image cache in each of the
compute nodes.  Using an image cache obviously reduces the boot time to a
minimum, since there is no transfer at all, therefore having a scheduler that
takes this into account is a need.  We obtained the best results when we
adapted the scheduler to take into account this cache, coupled with the usage
of BitTorrent as the image transfer method. Therefore, both solutions are
complementary: on the one hand we reduce the image transfer time when it is
needed, and on the other hand we profit from the cached images whenever
possible.

Taking into account those results, we think that there is room for future work
and improvements in the cloud scheduling algorithms so as to improve the boot
time for virtual machines. Cloud schedulers should be adapted to be cache-aware,
implementing at the same time policies that would ensure a compromise between
a fast boot time (i.e. the usage of a node with an image cached) and a fair
utilization of the resources (i.e. not constricting all request to be
scheduled only in one node).

On the other hand and taking into account the fact that users tend to request
images comprised in an small set of images (as shown in Figure~\ref{fig:pop}
and explained in Section~\ref{sec:distribution:prefetch}) we think that the
usage of popularity based distribution algorithms (so that the most used images
are available in the hosts) together with the cache aware scheduling would
introduce remarkable improvements in the deployment times. In this regard,
cloud monitoring \cite{Aceto2012} plays a key role, since one of the premises
for doing a proper pre-fetching is proper monitoring so as to get proper
metrics to evaluate if an image needs to be deployed or not.

\section*{Acknowledgements}

The authors acknowledge the financial support from the European Commission (via EGI-InSPIRE
Grant Contract number RI-261323).

The authors want also to thank the IFCA Advanced Computing and e-Science Group.

\bibliographystyle{elsarticle-num}
\bibliography{references}

\end{document}